\documentclass[aps,preprint,nofootinbib]{revtex4}%
\usepackage{eurosym}
\usepackage{xcolor} %%% Agregue este para marcar
\usepackage{amsfonts}
\usepackage{amsmath}
\usepackage{amssymb}
\usepackage{float}
\usepackage{graphicx}
\usepackage[utf8]{inputenc}%
\providecommand{\U}[1]{\protect\rule{.1in}{.1in}}
\DeclareMathOperator{\sech}{sech}
\DeclareMathOperator{\csch}{csch}
\begin{document}
\title{Scalar probes on wormholes in Lovelock theories with unique vacuum}
\author{Octavio Fierro$^{1}$, Daniela Narbona$^{2}$, Julio Oliva$^{2}$, Constanza
Quijada$^{2}$, Guillermo Rubilar$^{2}$}
\affiliation{$^{1}$Facultad de Ingeniería, Arquitectura y Diseño, Universidad San Sebastián, Lientur 1457, Concepción 4080871, Chile}
\affiliation{$^{2}$Departamento de F\'{\i}sica, Universidad de Concepci\'{o}n, Casilla
160-C, Concepci\'{o}n, Chile.}

\begin{abstract}
In this paper we construct new wormhole solutions of Lovelock theories in vacuum, when the coupling constants are such that all the maximally symmetric solutions coincide, extending to arbitrary dimensions wormhole solutions previously known in the Chern-Simons case. Like the latter, the wormholes are characterized by an integration constant $\rho_0$ that controls the contribution to the energy content from one of the boundaries. Then, we study the effects of the constant $\rho_0$ on the spectrum of a massive, (non)minimally coupled scalar probes, with Dirichlet boundary conditions at both asymptotic regions. As a result, a deformed Breitenlohner-Freedman bound emerges, which is sensitive to the value of $\rho_{0}$. The scalar spectra are numerically obtained in detail in dimension five, and in such dimension we also present a new family of wormhole geometries for the Einstein-Gauss-Bonnet theory with a unique vacuum. The new geometries are constructed via a double analytic continuation of a wormhole previously reported in the literature, but now the constant $\rho_0$ appears in the centrifugal terms of the equations for the geodesic and scalar probes. The mass of these configurations vanishes nontrivially, since the contributions to the mass integral from each boundary are nonvanishing, but only differ in sign, providing a new example of a spacetime having ``mass without mass."
\end{abstract}
\maketitle

\section{Introduction}

In the realm of four dimensional general relativity (GR), it is a difficult task to
construct asymptotically flat, spherically symmetric wormhole geometries since
they generally require violations of different energy conditions
\cite{Visser:1995cc}. In particular the averaged null energy condition must be
circumvented for a wormhole to exist if the throat is to provide a shorter
path connecting points of each asymptotic region through the bulk. The
obstructions can be avoided, for instance, by going beyond GR (see, e.g.,
\cite{Lobo:2012ai}, \cite{Bueno:2018uoy} and references therein), by the
inclusion of a NUT charge \cite{Ayon-Beato:2015eca}, or by constructing
wormholes with long throats supported by Casimir energy
\cite{Maldacena:2018gjk}. The inclusion of a negative cosmological constant
allowed to construct asymptotically AdS wormholes which in the radial
direction are foliated by warped AdS spacetimes and are devoid of closed
timelike curves \cite{Anabalon:2018rzq}, which possess a noncontratible
$S^{1}$, and contain a family of BPS configurations \cite{Anabalon:2020loe}.

The general features that a static metric must fulfill in order to describe a
traversable, asymptotically flat wormhole were studied in the seminal papers
\cite{Ellis:1973yv}, \cite{Morris:1988cz}, and \cite{Morris:1988tu}, and the
metrics considered in such references have been used as toy models to study
the propagation of probe fields on spacetimes with wormhole topology. For
example, the propagation of scalar and electromagnetic waves may correctly
lead to the interpretation of the wormhole geometry as an extended particle
\cite{Clement:1982ej}. Transmission and reflection coefficients for
asymptotically flat, ultrastatic wormholes in 2+1 and 3+1 dimensions have been
studied in Refs. \cite{Kar:1994ty}, \cite{Kar:1995jz}, and
\cite{Taylor:2014vsa}, featuring resonances for particular values of the
wormhole parameters, and leading to almost reflectionless effective
potentials\footnote{There is also an extensive literature on Euclidean
wormholes (see, e.g., \cite{Hebecker:2018ofv}) leading to instantons and on the use of scalar probes to test the 
stability of the configuration (see, e.g., \cite{Betzios:2017krj},  \cite{Jonas:2023qle} and references
therein). It has been recently established that geometries with such properties can be embedded in M-theory \cite{AnabalonGuarino}. In this paper we will be interested in traversable, Lorentzian
wormholes.}. In Ref. \cite{Taylor:2014vsa} it was also shown that a
nonminimally coupled scalar field may lead to an instability since the
effective Schr\"{o}dinger potential for the perturbation turns out to be
negative definite. Even more, when the wormhole is asymptotically flat in both
asymptotic regions, outgoing boundary conditions at both infinities lead to a
quasinormal spectrum which can mimic that of a black hole of mass $M$ for a
given wormhole mass $M^{-1}$ \cite{Taylor:2014vsa} (see also the recent works \cite{Khoo:2024yeh,Alfaro:2024tdr} and references therein). By imposing boundary conditions on the asymptotic AdS boundary as well as at
the throat, a particular family of smooth wormholes in Einstein's theory
coupled to a nonlinear electrodynamics has also been shown to be stable
provided an effective Breitenlohner-Freedman bound \cite{BF1, BF2, MT1} is
fulfilled for the probe scalar field \cite{Kim:2018ang}.
\bigskip

Asymptotically AdS wormholes in theories with quadratic terms in the curvature do exist
in five dimensions \cite{DOT1}, in vacuum. The action for the Einstein-Gauss-Bonnet
theory in five dimensions reads%
\begin{equation}
I=\frac{1}{16\pi G}\int\sqrt{-g}\left(  R-2\Lambda+\alpha\left(  R^{2}%
-4R_{\mu\nu}R^{\mu\nu}+R_{\alpha\beta\gamma\delta}R^{\alpha\beta\gamma\delta
}\right)  \right)  d^{5}x,
\end{equation}
where $\alpha$ is a coupling constant with mass dimension equal to $-2$. When
formulated in first order, the local Lorentz invariance of the theory is
enlarged to a local (A)dS group at the point $\Lambda\alpha=-3/4$ (see, e.g.,
\cite{MokhtarZanelliBook} and references therein). In this case, the Lagrangian
can be written as a Chern-Simons form where the spin connection as well as the
vielbein transform as components of an (A)dS gauge connection. This structure
can be extended to higher odd-dimensions as well as to dimension three, but in
contrast to the latter, the higher dimensional case does describe a theory
with local degrees of freedom in the bulk. As shown in \cite{Banados:1993ur},
all these theories contain generalizations of the static $2+1$-dimensional BTZ
black hole \cite{BTZ1}-\cite{BTZ2}, characterized by an integration constant
which can be identified with the mass. In \cite{DOT1} it was shown that this theory actually
admits a larger family of static solutions, including wormhole geometries in
vacuum with two asymptotically locally AdS$_{2n+1}$ regions for $n\geq2$. In
five dimensions, the line element of these solutions is given by%
\begin{equation}
ds^{2}=l^{2}\left[  -\cosh^{2}\left(  \rho-\rho_{0}\right)  dt^{2}+d\rho
^{2}+\cosh^{2}\rho\left(  d\phi^{2}+d\Sigma_{2}^{2}\right)  \right]
,\label{wormhole}%
\end{equation}
where $-\infty<t<\infty$, $-\infty<\rho<\infty$, $0\leq\phi<2\pi$, identified, and
$d\Sigma_{2}$ stands for the line-element of an Euclidean $2$-dimensional
manifold, whose local geometry is that of the hyperbolic space with radius
$3^{-1/2}$, globally equivalent to the quotient of the hyperbolic space by a
Fuchsian group, i.e., $\Sigma_{2}$ is homeomorphic to the quotient of $H_{2}$
by a freely acting, discrete subgroup of $SO(2,1)$. For any local purpose one
can consider complex projective coordinates such that%
\begin{equation}
d\Sigma_{2}^{2}=\frac{1}{3}\left(  1-\frac{z\bar{z}}{4}\right)  ^{-2}%
dzd\bar{z}.
\end{equation}

The spacetime (\ref{wormhole}) represents a wormhole geometry with a throat
located at $\rho=0$, connecting two asymptotically locally AdS spacetimes with
curvature radius $l^{2}=4\alpha=-3/\Lambda$. The
constant $\rho_{0}$ is an arbitrary integration constant and determines the
apparent mass of the wormhole as seen by an asymptotic observer at a given
asymptotic region \cite{DOT1}-\cite{DOT2}. The presence of $\rho_{0}\neq0$,
creates a region which has interesting properties for particles with angular
momentum, probing these geometries. As shown in \cite{DOT2}, the effective
potential for the geodesics has two contributions, one of which is due to the
angular momentum of the particle around the $S^{1}$ factor parametrized by the
coordinate $\phi$ in (\ref{wormhole}). This centrifugal contribution always
points outward the throat located at $\rho=0$, while the remaining
contribution flips its direction at the surface $\rho=\rho_{0}$ and always
points toward it. Therefore, the gravitational pull acting on a particle can
be balanced by the centrifugal contribution only if the particle is in one of
the regions $-\infty<\rho<0$ or $\rho_{0}<\rho<+\infty$, for positive $\rho_{0}$. Geodesic probes turn
out to be expelled from the region $0<\rho<\rho_{0}$. And for probe strings
propagating in this background the surface $\rho=\rho_{0}/2$ defines the
turning point of strings with both ends attached to the same asymptotic region
\cite{Ali:2009ky}.

% The main goal of the present work is to explore the effect of $\rho_{0}\neq0 $
% on the propagation of massive and massless probe scalars, with both minimal
% and non-minimal coupling with the curvature, in arbitrary dimension. We will
% be interested in the normal frequencies of the scalars obtained by imposing
% reflective boundary conditions at the AdS boundaries.

In arbitrary odd dimensions, $d=2n+1$, these wormholes can be embedded in Lovelock theories at the Chern-Simons point, at which the Lagrangian can be written as a Chern-Simons form for the $SO(2n,2)$ group. In this work, first we show that in even dimension, $d=2n$, the wormhole geometry can also be embedded in Lovelock theory with a unique vacuum, when the maximum power in
the curvature is present. Therefore we would have identified a sensible
gravity theory in every dimension for which the wormhole geometry is a solution. Then
we will study a minimally coupled massive scalar field on the wormhole
geometry (\ref{wormhole}), revisiting the case with $\rho_{0}=0$ which can be
solved analytically (see \cite{Correa:2008nq}). Following a complementary
approach to that in Ref. \cite{Correa:2008nq}, by formulating the problem
in a Schr\"{o}dinger form, we show that the effective potential for the
scalar probe corresponds to a Rosen-Morse potential, which explains the
integrability of the spectrum \cite{Cooper:1994eh}. Remarkably, we find that
also in the limit $\rho_{0}\rightarrow\infty$ the spectrum can be obtained in
a closed form as well, which would be particularly useful for obtaining the normal
frequencies of the scalar on wormholes with large $\rho_{0}$. Then, we will
numerically solve the spectrum for normal frequencies of the scalar field for
finite, nonvanishing values of $\rho_{0}$ in dimension five, connecting the two exactly
solvable models. We also show that there is an effective
Breitenlohner-Freedman mass for the scalar probe, which depends on
$\rho_{0}$. In the last section, via a double Wick rotation, we construct a new family of wormholes, which maintain the properties of the asymptotic regions. We also show that the propagation of a scalar probe on the new family of wormholes, is equivalent to the propagation on the former geometries, by performing the double Wick rotation at the level of the quantum numbers, namely by setting $\omega\rightarrow in$ and $n\rightarrow i\omega$. Finally we obtain the mass of this new wormhole geometry.

\section{Embedding the wormhole in Lovelock theories in even dimensions}

Before analysing the propagation of a scalar probe on the wormhole geometry in
arbitrary dimensions, below we identify a Lovelock theory which admits the wormhole as a
vacuum solution, in even dimension. As mentioned before, in odd dimensions it
is enough to consider Lovelock theory with the couplings related in such a
manner that the action can be written as a Chern-Simons form for the AdS
group. Since Lovelock theories with generic couplings fulfill a Birkhoff's
theorem (see \cite{Zegers:2005vx,Ray:2015ava}), we will have to consider some relation
between the couplings in order to by-pass such uniqueness result. For concreteness, let us focus on the Lovelock theory
that has a unique maximally symmetric solution and contains all the possible
powers of the curvature allowed in dimension $d$, i.e., we consider all the
Lovelock terms of the form $\mathcal{R}^{k}$ with $k\leq [(d-1)/2]$. The field
equations of such theory can be written in a very compact manner
\begin{equation}\label{LUV}
E_{B}^{A}:=\delta_{BD_{1}...D_{2k}}^{AC_{1}...C_{2k}}\bar{R}_{C_{1}C_{2}}^{D_{1}D_{2}}...\bar{R}_{C_{2k-1}C_{2k}}^{D_{2k-1}D_{2k}}=0\ ,
\end{equation}
where $\bar{R}_{C_{1}C_{2}}^{D_{1}D_{2}}:=R_{C_{1}C_{2}}^{D_{1}D_{2}}%
+l^{-2}\delta_{C_{1}C_{2}}^{D_{1}D_{2}}$. Here $l$ is the curvature radius of
the unique, maximally symmetric, AdS solution of the theory. Notice that all
allowed powers of the curvature appear in the field equations. It is easy to
see that for the wormhole geometry%
\begin{equation}
ds^{2}=l^{2}\left[  -\cosh^{2}\left(  \rho-\rho_{0}\right)  dt^{2}+d\rho
^{2}+\cosh^{2}\rho d\Sigma_{d-2}^{2}\right]\ ,
\end{equation}
the components of the shifted curvature $\bar{R}_{CD}^{AB}$ are%
\begin{align}
\bar{R}_{\ \ \ t\rho}^{t\rho}  &  =\bar{R}_{\ \ \ \rho j}^{\rho i}=0,\ \bar
{R}_{\ \ \ tj}^{ti}=\frac{\left(1  -\tanh\rho\tanh\left(  \rho-\rho_{0}\right)
\right)  }{l^{2}}\delta_{j}^{i}\ ,\\
\bar{R}_{\ \ \ kl}^{ij}  &  =\tilde{R}_{\ \ \ kl}^{ij}+\delta_{kl}^{ij}\ ,
\end{align}
where Latin indices $\{i,j,k,l\}$ are coordinate indices on the manifold
$\Sigma_{d-2}$, which has an intrinsic Riemann tensor $\tilde{R}%
_{\ \ \ kl}^{ij}$. As explained in \cite{DOT1}, in odd dimension $d=2k+1$ the field
equations imply a single scalar constraint on the Euclidean manifold
$\Sigma_{d-2}$
\begin{equation}
\delta_{j_{1}...j_{2k-2}}^{i_{1}...i_{2k-2}}\left(  \tilde{R}_{\ \ \ i_{1}%
i_{2}}^{j_{1}j_{2}}+\delta_{i_{1}i_{2}}^{j_{1}j_{2}}\right)  ...\left(
\tilde{R}_{\ \ \ i_{2k-3}i_{2k-2}}^{j_{2k-3}j_{2k-2}}+\delta_{i_{2k-3}%
i_{2k-2}}^{j_{2k-3}j_{2k-2}}\right)  =0\ ,
\end{equation}
which is solved for example by the manifold $S^{1}\times H_{d-3}$, with
$H_{d-3}$ with a suitable radius.

Here we are interested in the embedding of the wormhole geometry in a Lovelock
theory with unique vacuum, in even dimension with $d=2k+2$. In this case the
field equations \eqref{LUV} reduce to a tensor and a scalar constraint on the manifold
$\Sigma_{d-2}$, which respectively read%
\begin{align}\label{tensorconst}
\delta_{lj_{1}...j_{2k}}^{ki_{1}...i_{2k}}\left(  \tilde{R}_{\ \ \ i_{1}i_{2}%
}^{j_{1}j_{2}}+\delta_{i_{1}i_{2}}^{j_{1}j_{2}}\right)  ...\left(  \tilde
{R}_{\ \ \ i_{2k-3}i_{2k-2}}^{j_{2k-3}j_{2k-2}}+\delta_{i_{2k-3}i_{2k-2}%
}^{j_{2k-3}j_{2k-2}}\right)   &  =0\ ,\\
\delta_{j_{1}...j_{2k}}^{i_{1}...i_{2k}}\left(  \tilde{R}_{\ \ \ i_{1}i_{2}%
}^{j_{1}j_{2}}+\delta_{i_{1}i_{2}}^{j_{1}j_{2}}\right)  ...\left(  \tilde
{R}_{\ \ \ i_{2k-1}i_{2k}}^{j_{2k-1}j_{2k}}+\delta_{i_{2k-1}i_{2k}}%
^{j_{2k-1}j_{2k}}\right)   &  =0\ .\label{scalarconst}
\end{align}
To fix ideas, if one considers these constraints in dimension six, they respectively reduce to that of
an Einstein manifold $\tilde{R}_{\ j}^{i}=-3\delta_{j}^{i}$ with a constant
Kretchman scalar $\tilde{R}_{ijkl}\tilde{R}^{ijkl}=36$. In consequence, provided we fulfill these conditions, we would have found a new wormhole solution of Lovelock theory with a unique vacuum in even dimensions. A simple inspection of these equations show that one can solve them by considering suitable products of constant curvature spacetimes, or even products of homogeneous geometries (see \cite{Anabalon:2011bw,Matulich:2011ct}).

\section{Normal modes of the scalar probe}
The scalar probe
\begin{equation}\label{KG}
\left(  \square-m^{2}\right)  \Phi\left(  x^{\mu}\right)  =0 \ ,
\end{equation}
on the wormhole geometry
\begin{equation}
ds^{2}=l^{2}\left[  -\cosh^{2}\left(  \rho-\rho_{0}\right)  dt^{2}+d\rho
^{2}+\cosh^{2}\rho\left(  d\phi^{2}+d\Sigma_{d-3}^{2}\right)  \right]
,\label{wormholed}%
\end{equation}
can be separated as
\begin{equation}
\Phi\left(  x\right)  =\mathcal{R}\left(  \sum_{n=-\infty}^{\infty}\int
d\omega\ R_{\omega,n}\left(  \rho\right)  \ e^{-i\omega t+in\phi}\right)
\ ,\label{separacion}%
\end{equation}
and hereafter we will drop the dependence of $R_{\omega,n}(\rho)$ on $\omega$ and $n$. Notice that we have assumed the scalar probe to be independent of the coordinates parametrizing the manifold $\Sigma_{d-3}$. Since the equation is linear, one finds decoupled, second order,
linear ODEs for each mode, leading to
\begin{align}
0 & =\partial_\rho\left[\cosh(\rho-\rho_0)\cosh^{d-2}(\rho)\partial_\rho R(\rho)\right]+\nonumber\\
& +\cosh(\rho-\rho_0)\cosh^{d-2}(\rho)\left[\frac{\omega^2}{\cosh^2(\rho-\rho_0)}-\frac{n^2}{\cosh^2(\rho-\rho_0)}-m^2l^2\right]R(\rho) \label{ecuRrho}
\end{align}

% \begin{align}
% 0 &  =\cosh^{2}\left(  \rho\right)  \frac{d^{2}R}{d\rho^{2}}+\left(
% 3\cosh\left(  \rho\right)  \sinh\left(  \rho\right)  +\frac{\sinh\left(
% \rho-\rho_{0}\right)  \cosh^{2} \left(  \rho\right)  }{\cosh\left(  \rho
% -\rho_{0}\right)  }\right)  \frac{dR}{d\rho}\nonumber\\
% &  -\left(  \cosh^{2}\left(  \rho\right)  m^{2}+n^{2}-\frac{\omega^{2}%
% \cosh^{2}\left(  \rho\right)  }{\cosh^{2}\left(  \rho-\rho_{0}\right)
% }\right)  R\left(  \rho\right)  .\label{ecuRrho}%
% \end{align}
Had we considered dependence on the coordinates of $\Sigma_{d-3}$, by including in \eqref{separacion} an eigenfunction of the Laplace operator on $\Sigma_{d-3}$ denoted by $Y_k\left(\sigma_{d-3}\right)$, the Eq. \eqref{ecuRrho} would have acquired a modification of the form $n^2\rightarrow n^2+k^2$, where $-k^2$ is the eigenvalue of the Laplace operator on $\Sigma_{d-3}$.

Introducing the inversion $z=(1-\tanh\rho)/{2}$, we map $-\infty<\rho<+\infty$
to $1>z>0$, and obtain the asymptotic behavior for the radial dependence of
the scalar field:
\begin{equation}
R\left(  z\right)  \rightarrow c_{1}z^{\Delta_{+}}\left[  1+\mathcal{O}%
(z)\right]  +c_{2}z^{\Delta_{-}}\left[  1+\mathcal{O}(z)\right]  \text{ as
}z\rightarrow0,\label{asymp}%
\end{equation}
and%
\begin{equation}
R\left(  z\right)  \rightarrow d_{1}\left(  1-z\right)  ^{\Delta_{+}}\left[
1+\mathcal{O}(1-z)\right]  +d_{2}\left(  1-z\right)  ^{\Delta_{-}}\left[
1+\mathcal{O}(1-z)\right]  \text{ as }z\rightarrow1,\label{2}%
\end{equation}
with%
\begin{equation}
\Delta_{\pm}=\left(\frac{d-1}{4}\right)\pm\frac{1}{2}\sqrt{m^{2}+\left(\frac{d-1}{2}\right)^2}\ ,
\end{equation}
$c_{1,2}$ and $d_{1,2}$ being integration constants.

Even though scalar fields on asymptotically AdS spacetimes can have negative
$m^{2}$, we will focus on the case $m^{2}\geq0$ (for a thorough analysis of
the properties of a scalar field on the particular case of the wormhole with
$\rho_{0}=0$ and the deformed Breitenlohner-Freedman bound, see
\cite{Correa:2008nq}). Reflective boundary conditions require the field to
vanish at infinity and therefore only the $\Delta_{+}$ branch in (\ref{asymp})
and (\ref{2}) is allowed. Under these boundary conditions the equation
(\ref{ecuRrho}) defines a Sturm-Liouville problem which can be manifestly seen
by transforming the radial problem (\ref{ecuRrho}) into a Schr\"{o}dinger-like
equation.

In order to simplify the presentation, here after, let us fix the dimension to $d=5$. Equation \eqref{ecuRrho} can be written as%
\begin{equation}
-\frac{d^{2}u}{d\bar{\rho}^{2}}+U(\bar{\rho})u=\omega^{2}%
u\ ,\label{schrogeneral}%
\end{equation}
where%
\begin{equation}
u=R\cosh^{3/2}(\rho)\quad\text{and}\quad\rho=\rho_{0}+\ln\left(  \tan\left(
\frac{\bar{\rho}}{2}\right)  \right)  \ .\label{cambiodevariables}%
\end{equation}
The radial coordinate $\bar{\rho}$ works as a \textquotedblleft
tortoise"-like coordinate in the sense that in terms of $(t,\bar\rho)$ the
two-dimensional part of the metric (\ref{wormhole}) is manifestly conformally
flat. The coordinate $\bar\rho$ connects both asymptotically AdS regions
$0<\bar{\rho}<\pi$. The effective Schr\"{o}dinger potential takes the form%
\begin{align}
U(\bar{\rho}) &  =\frac{\left(  4n^{2}-3\right)  }{4}\left(  \frac{\cos
(\bar{\rho})\cosh(\rho_{0})-\sinh(\rho_{0})}{\cos(\bar{\rho})-\coth(\rho_{0}%
)}\right)  ^{2}-\left(  2n^{2}-3\right)  \left(  \frac{\cos(\bar{\rho}%
)\cosh(\rho_{0})-\sinh(\rho_{0})}{\cos(\bar{\rho})\sech(\rho_{0}%
)-\csch(\rho_{0})}\right) \nonumber\\
&  \ \ +\frac{\cosh^{2}(\rho_{0})\left(  4n^{2}-9\right)  }{4}+\frac{\left(
4m^{2}+15\right)  }{4\sinh^{2}(\bar{\rho})}\, .
\end{align}
\begin{figure}[h]
\includegraphics[scale=0.4]{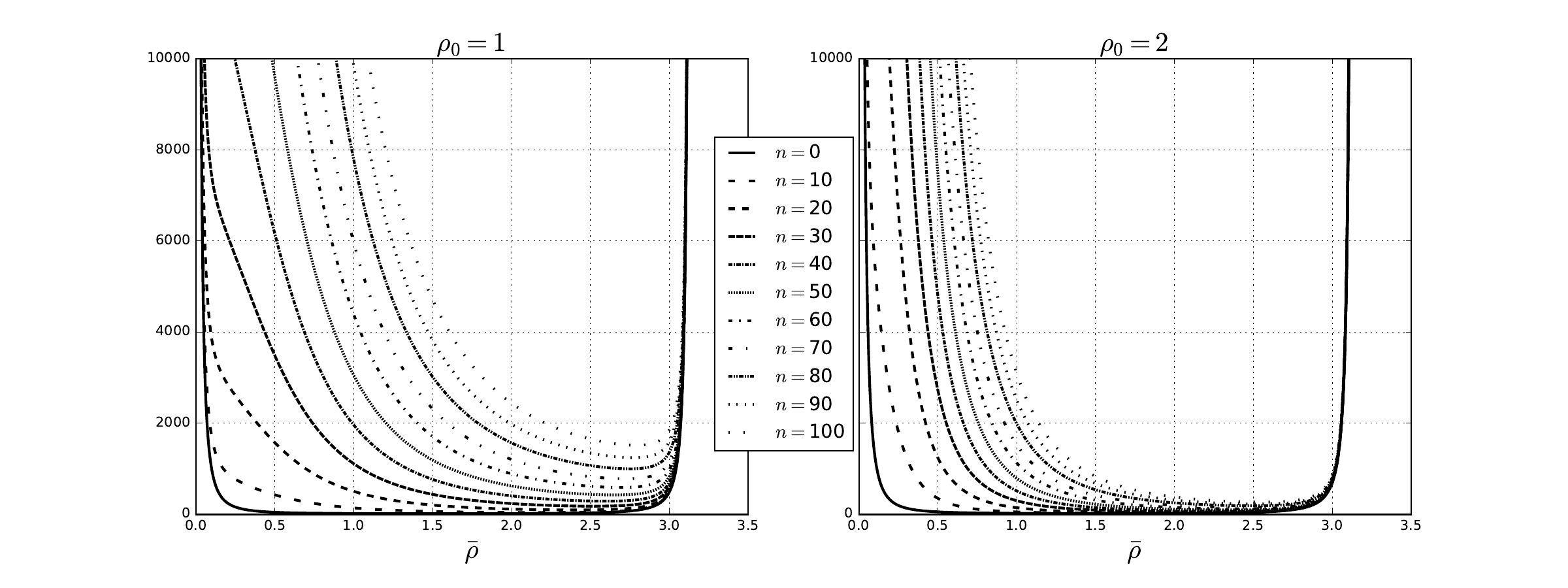} \caption{Effective potential
for the radial dependence of the scalar probe, for different values of the
parameters. As expected in the presence of a negative cosmological constant,
the potential diverges at the boundaries $\bar\rho=0,\pi$.}%
\label{potentials}%
\end{figure}
This potential $U(\bar{\rho})$ (depicted in Fig. 1),
parametrically depends on the wormhole integration constant $\rho_{0}$, and
for arbitrary values of the latter, the equation cannot be solved
analytically. Even though we will solve the equation numerically to find the
normal modes of the scalar probe on the wormhole, it is interesting to note that
there are two particular values of $\rho_{0}$ for which the potential is shape
invariant \cite{Cooper:1994eh} and \eqref{schrogeneral} can be solved
analytically. Those values are $\rho_{0}=0$ and $\rho_{0}\rightarrow+\infty$,
and the corresponding potentials $U$ are given by
\begin{align}
U^{\left(  0\right)  }(\bar{\rho}) &  :=\frac{15}{4}\frac{1}{\sin^{2}\bar
{\rho}}+\frac{m^{2}}{\sin^{2}\bar{\rho}}-\frac{9}{4}+n^{2}+\mathcal{O}\left(
\rho_{0}\right)  \ ,\\
U^{\left(  \infty\right)  }(\bar{\rho}) &  :=-\frac{1}{4}\frac{6\cos\bar{\rho
}-4m^{2}-9}{\sin^{2}\bar{\rho}}+\mathcal{O}\left(  e^{-2\rho_{0}}\right)
\ .\label{infschr}%
\end{align}

\bigskip

The leading term of $U^{\left(  0\right)  }$ leads to an effective quantum
mechanical problem with a Rosen-Morse potential, which can be solved
analytically and with energies and bound states given by
\begin{align}
\omega_{\left(  0\right)  ,p}^{2} &  =\left(  \frac{1}{2}+\sqrt{4+m^{2}%
}+p\right)  ^{2}+n^{2}-\frac{9}{4}\ ,\\
u_{p}^{\left(  0\right)  }\left(  \bar{\rho}\right)   &  =A_{p}^{\left(
0\right)  }\left(  \sin\bar{\rho}\right)  ^{s+p}P_{p}^{\left(
-s-p,-s-p\right)  }\left(  i\cot\left(  \bar{\rho}\right)  \right)  \ ,
\end{align}
with $s=\sqrt{4+m^{2}}+{1}/{2}$, $A_{p}^{\left(  0\right)  }$ an arbitrary
integration constant that can be fixed by normalization, and $p=0,1,2,3,...$
as the mode number. For the particular value $\rho_{0}=0$ the wormhole
acquires a reflection symmetry with respect to the throat. $P_a^{\ b}$ are the
corresponding Jacobi polynomials.

\bigskip

Remarkably, for $\rho_{0}\rightarrow\infty$, the Schr\"{o}dinger problem in
(\ref{infschr}) can also be integrated analytically since it corresponds to a
Scarf potential, and leads to the following frequencies and eigenfunctions
\begin{align}
\omega_{\left(  \infty\right)  ,p}^{2} &  =\left(  A+p\right)  ^{2}%
\text{\ ,}\\
u_{p}^{\left(  \infty\right)  }\left(  \bar{\rho}\right)   &  =A_{p}^{\left(
\infty\right)  }\left(  1-\cos\left(  \bar{\rho}\right)  \right)  ^{\frac
{A-B}{2}}\left(  \cos\left(  \bar{\rho}\right)  +1\right)  ^{\frac{A+B}{2}%
}P_{p}^{\left(  A-B-\frac{1}{2},A+B-\frac{1}{2}\right)  }\left(  \cos\left(
\bar{\rho}\right)  \right)  \ ,
\end{align}
where the constant $A>B$ are given by%
\begin{align}
2A  & =1+\sqrt{5+2m^{2}+2\sqrt{\left(  m^{2}+1\right)  \left(  m^{2}+4\right)
}}\ ,\\
6B  & =\left(  5+2m^{2}-2\sqrt{\left(  m^{2}+1\right)  \left(  m^{2}+4\right)
}\right)  \sqrt{5+2m^{2}+2\sqrt{\left(  m^{2}+1\right)  \left(  m^{2}%
+4\right)  }}%
\end{align}
$A_{p}^{\left(  \infty\right)  }$ is an integration constant and $p=0,1,2,...$
, is again the mode number. It is worth to mention that in this case the
frequencies $\omega_{p}^{\left(  \infty\right)  }$ lead to an equispaced,
fully resonant spectrum. This result may be particularly relevant for
obtaining the spectrum of the scalar field on the wormhole, for large
$\rho_{0}$, by perturbative methods. The strict limit $\rho_{0}$ to infinity
can be taken after suitable regularization of the geometry \cite{DOT2},
leading to a wormhole that connects two different asymptotic regions. Since
such wormhole is not asymptotically locally AdS we left its analysis for the Appendix.

\bigskip

Here we are interested in the effects of a finite value of $\rho_{0}$ on the
propagation of a minimally coupled scalar field. Therefore, we are obligated
to integrate equation (\ref{schrogeneral}) numerically, with reflective
boundary conditions at both infinities which can be achieved only for a
countably infinity number of frequencies $\omega_{p}$. This is done by using a
variation of the so-called shooting method (see, for instance,
\cite{Newman2013}), which in our case consists in varying the value of
$\omega$ in \eqref{schrogeneral} and seeking those values which fulfill the
boundary condition at $\bar{\rho}=\pi$. More specifically, the Schr\"odinger
equation is recast as a first order ODE system, which is then integrated
``from left to right", i.e. from $\bar{\rho}=0$ to $\bar{\rho}=\pi$ for each
value of $\omega$. The integration is performed using the standard 4th order
Runge-Kutta method \cite{Newman2013}, which thus determines the value of the
solution at the right boundary, i.e. $u(\bar{\rho}=\pi,\omega)$. The later can
be considered as a function of $\omega$, for which the sought after values
$\omega_{p}$ are zeros of, in virtue of the right boundary condition
$u(\bar{\rho}=\pi,\omega_{p})$. After a solution for $\omega_{p}$ is found,
the value of $p$ is determined by counting the number of nodes of the function
$u(\bar{\rho},\omega_{p})$.

\section{Spectra for the minimally coupled scalar probe}

Since the problem for the normal modes with $\rho_{0}\neq0$ is a
Sturm-Liouville problem, with potential fixed by the angular momentum of the
field $n$, as well as $\rho_{0}$ and $m^{2}$, the eigenfunctions and
eigenfrequencies will both be labeled by an integer $p$. In all the plots
shown below, we have included small black bars near $\rho_{0}=0$ and $\rho
_{0}$ large, marking the analytic values for the frequencies obtained in those
cases. In what follows we plot some of the numerically obtained spectra, which
present interesting features as $\rho_{0}$ varies. \newline The frequencies of
the fundamental mode ($p=0$) are shown on the left panel of Fig. 2, while the
right panel shows a set of frequencies for the tenth overtone ($p=10$). While
the fundamental mode is a monotonic function of $\rho_{0}$, we observe that
the excited modes have a maximum frequency for a critical value of $\rho_{0}$
and then decay to a given value. As expected, the frequencies increase with
the angular momentum of the field, and consistently with the asymptotic
expression for $U^{(\infty)}$ in Eq. (\ref{infschr}), the frequencies
merge regardless the value of $n$.

Figure 3 shows the fundamental and ten excited modes for the ``s-wave" of the
scalar ($n=0$) on the left panel, while the frequencies for a spinning scalar
probe with $n=10$ have been plotted on the right-panel, both as a function of
$\rho_{0}$. It is particularly interesting to notice that the ``s-wave"
fundamental and excited frequencies remain almost constant regardless the
value of the integration constant $\rho_{0}$.

Finally, Fig. 4 shows the fundamental and first overtones for different values
of the angular momentum and the mass of the scalar. Even though the effective
Schr\"{o}dinger problem is one-dimensional, there are degeneracies due to
the fact the effective potential depends on $n$, therefore different
values of $n$ lead to different one-dimensional Sturm-Liouville problems.

\begin{figure}[h]
\includegraphics[scale=0.4]{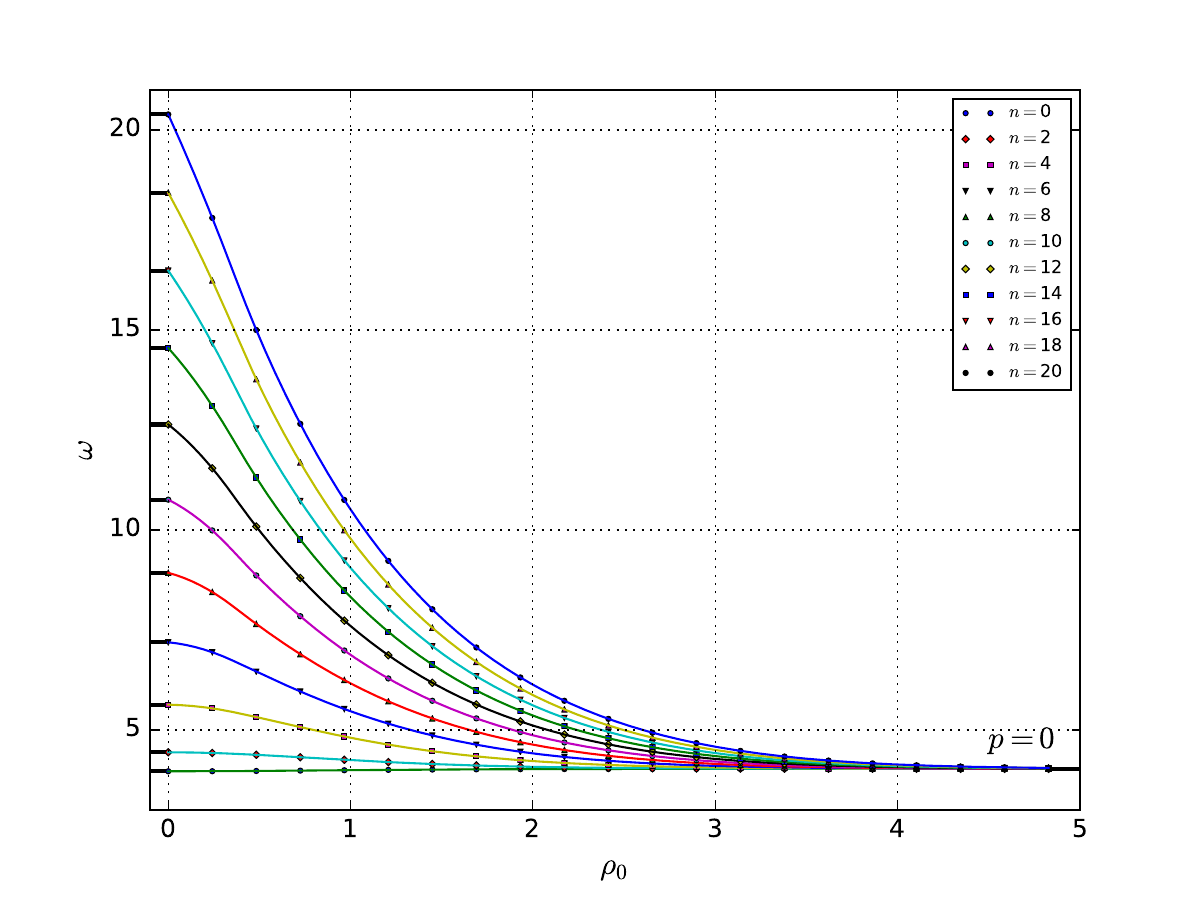}
\includegraphics[scale=0.4]{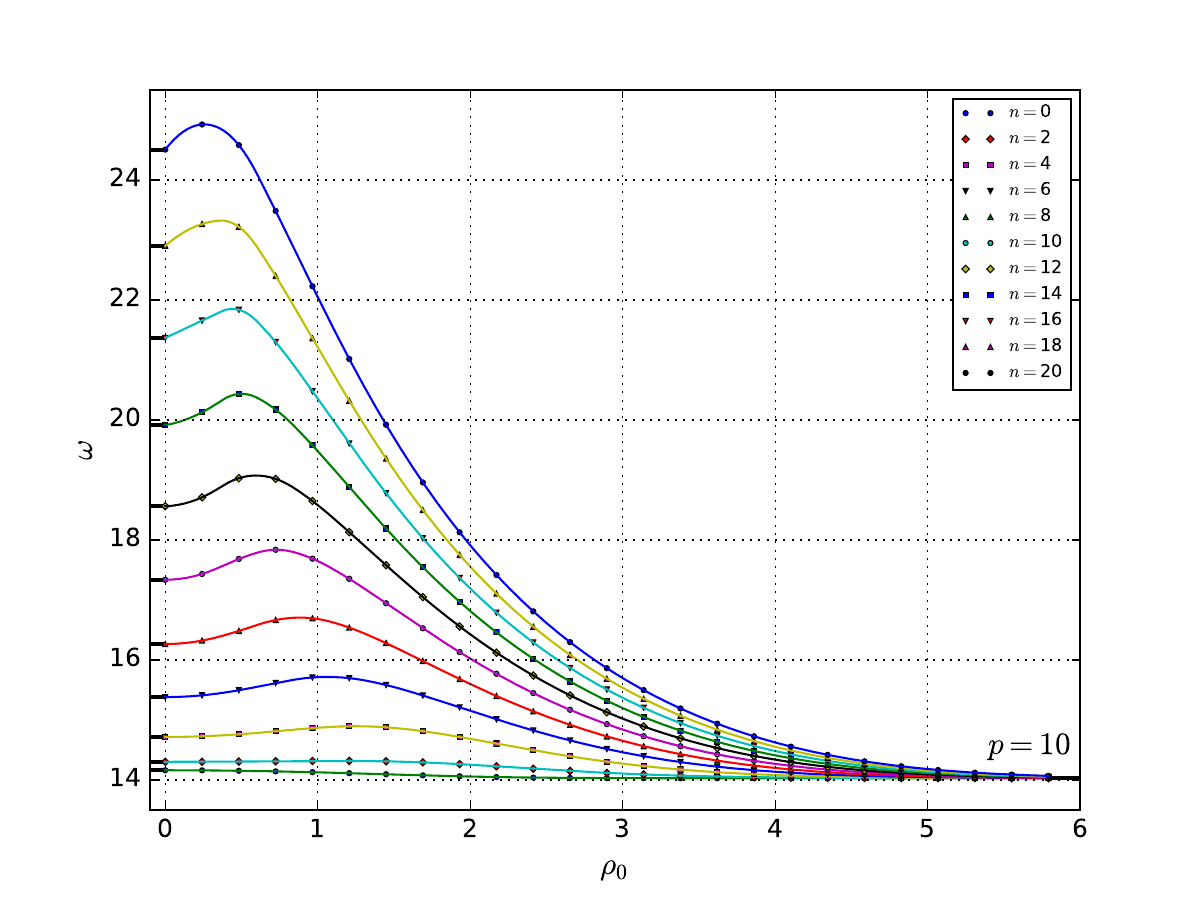} \caption{Spectra for the
fundamental mode $p=0$ (left-panel) and tenth overtone $p=10$ (right-panel) as
a function of $\rho_{0}$ for different even values of the angular momentum of
the scalar probe $n=0,2,4,...,10,$ for $m^{2}=10$. The fundamental mode has a
monotonic behavior with $\rho_{0}$ while excited frequencies present a
maximum for a given critical value of $\rho_{0}$ which depends on the angular
momentum of the field.}%
\label{p0p10xi0}%
\end{figure}

\begin{figure}[h]
\includegraphics[scale=0.4]{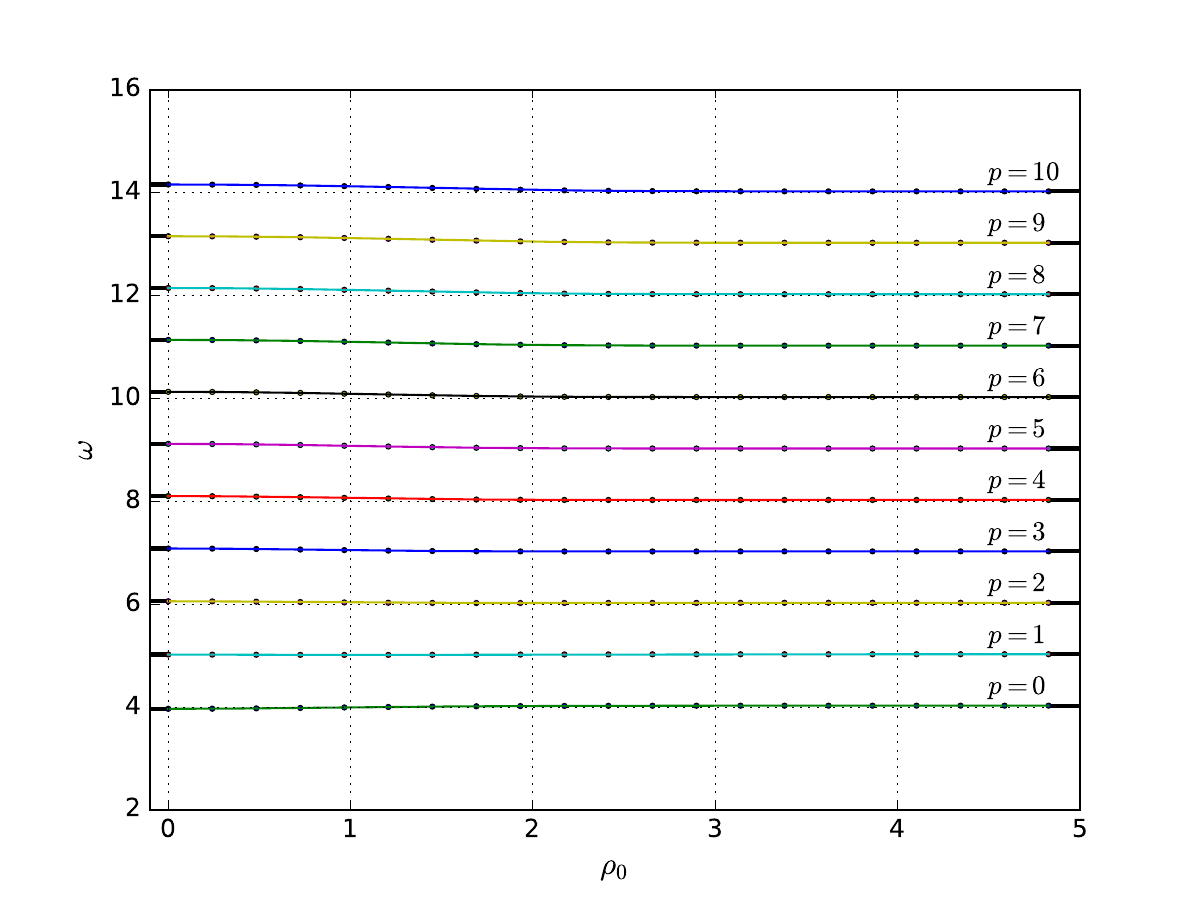}
\includegraphics[scale=0.4]{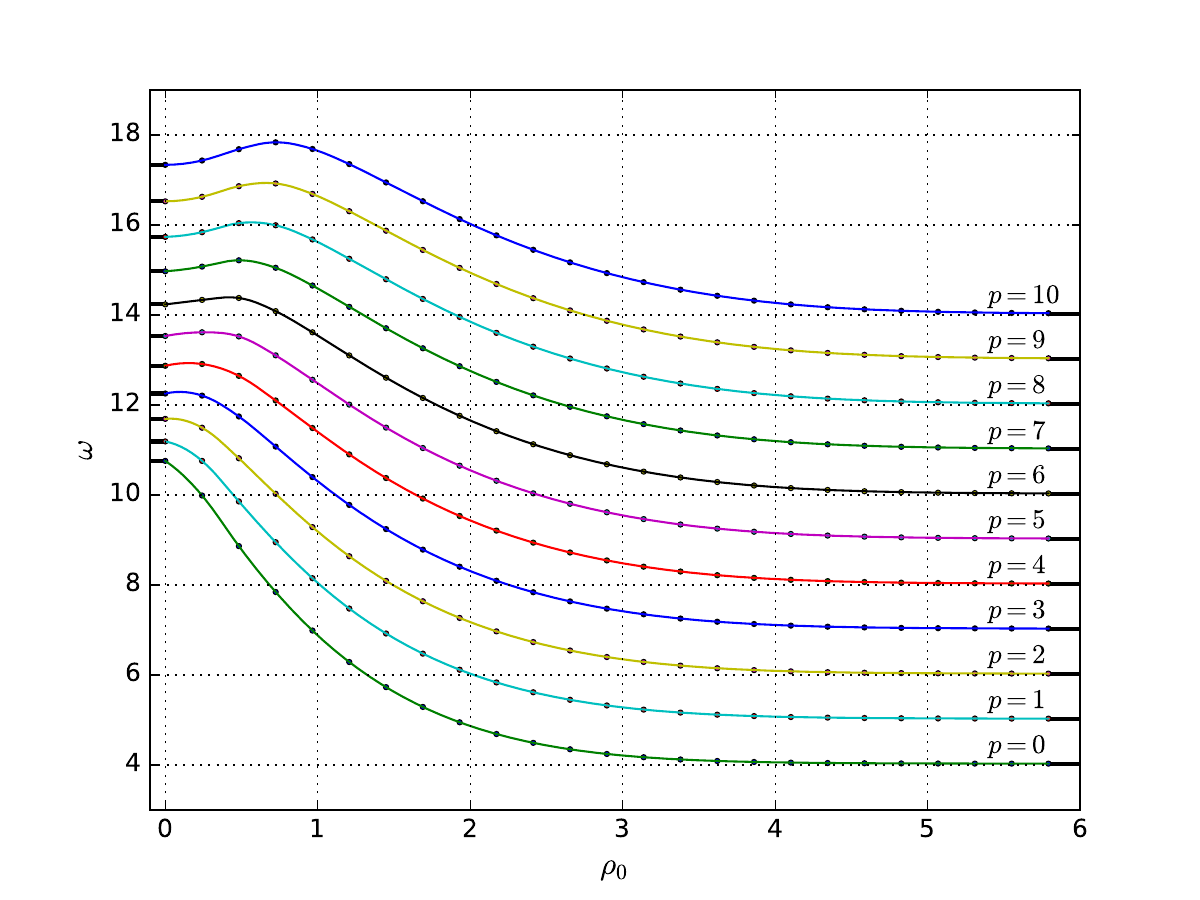} \caption{Spectra for
vanishing angular momentum $n=0$ (left-panel) and a spinning scalar probe with
$n=10$ (right-panel) as a function of $\rho_{0}$ for the fundamental mode and
the first 10 excited states ($m^{2}=10$).}%
\label{n0n10xi0}%
\end{figure}

\begin{figure}[h]
\includegraphics[scale=0.4]{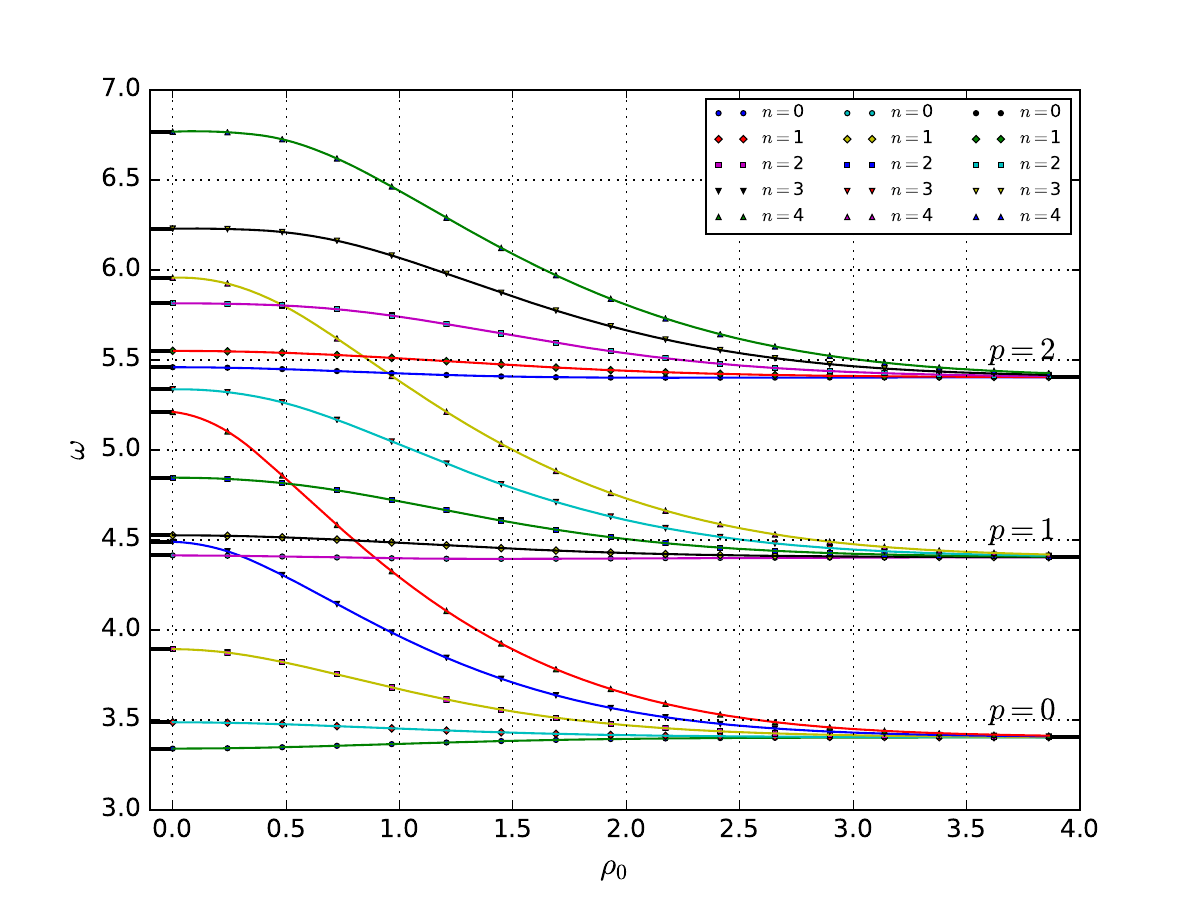}
\includegraphics[scale=0.4]{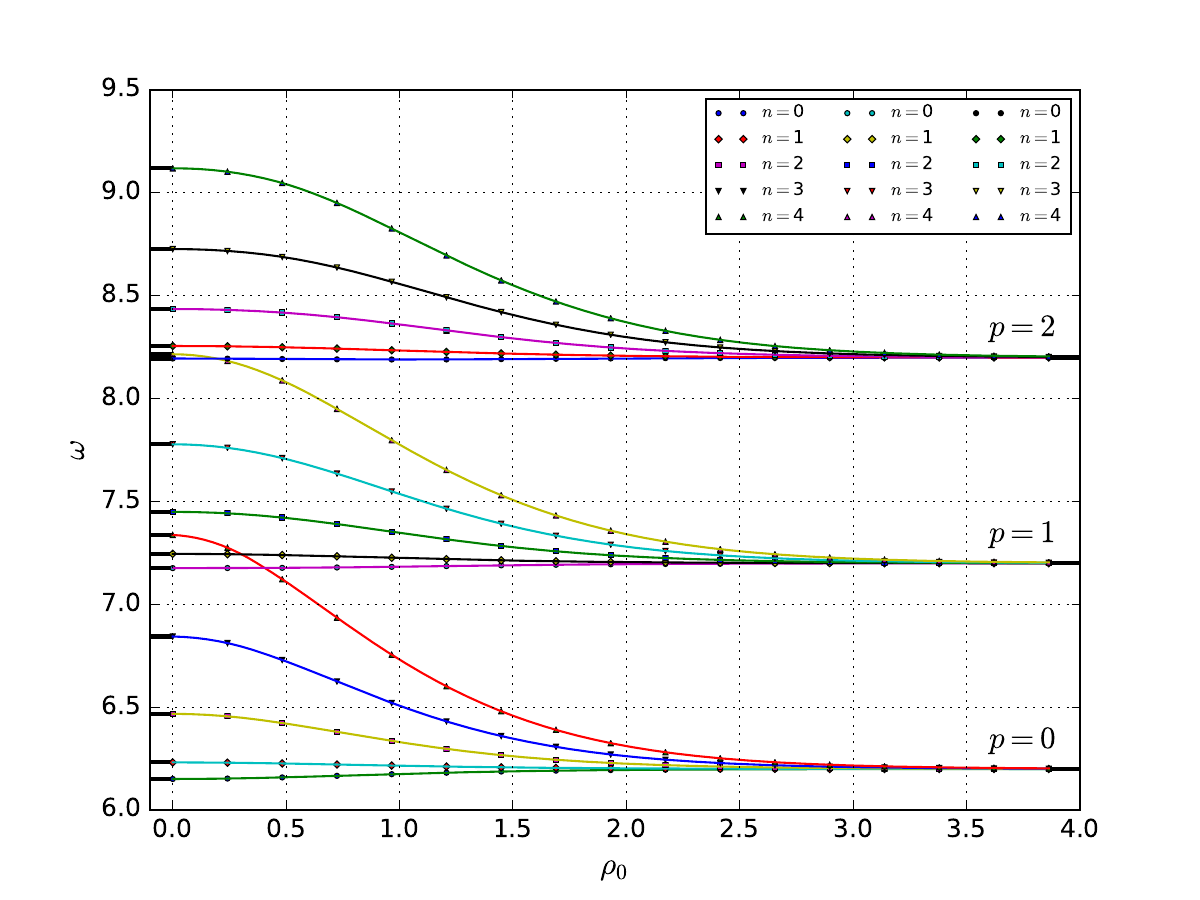} \caption{Fundamental and
first two overtones for the scalar probe with $n=0,1,2,3,4$, and $m^{2}=6$
(left-panel) and $m^{2}=30$ (right-panel). For a given value of $\rho_{0}$ the
same frequency can be obtained for different modes since the effective
potential depends explicitly on $n$.}%
\label{n0n10xi0}%
\end{figure}\newpage

\section{Scalars probes with nonminimal coupling}

As shown in \cite{Correa:2008nq}, for the case $\rho_{0}=0$ the equation for a
scalar nonminimally coupled with the scalar curvature can also be solved in
an analytic manner. This is remarkable since the Ricci scalar of the wormhole
background is a nontrivial function of the radial coordinate $\rho$, therefore including a nonminimal coupling with the scalar curvature do not stand for a shift in the mass. Here, we explore
the effect of a nonminimal coupling on the spectrum of the scalar for
$\rho_{0}\neq0$, still within the context of dimension five. The equation for the scalar in this case is given by%

\begin{equation}
\left(  \square-m^{2}-\xi R\right)  \Phi\left(  x^{\mu}\right)  =0\, .
\end{equation}

This scalar probe is invariant under local Weyl rescalings for $\xi=3/16$.
When $\rho_{0}$ vanishes, the equation for the radial dependence of the
nonminimally coupled scalar can be obtained from that of the minimally
coupled one by shifting the mass as well as the eigenvalues of the Laplace
operator on the manifold $\Sigma_{3}=S^{1}\times H_{2}/\Gamma$. For
nonvanishing $\rho_{0}$, one cannot rely on this shift, and one is forced to
numerically find the spectrum of normal modes. Using the separation
\eqref{separacion} and the change of variables \eqref{cambiodevariables}, one
leads to a Schr\"{o}dinger-like equation of the form \eqref{schrogeneral} with
an intricate potential given by%

\begin{align}
U_{\xi}(\bar{\rho}) &  =\frac{\left(  4n^{2}-3\right)  }{4}\left(  \frac
{\cos(\bar{\rho})\cosh(\rho_{0})-\sinh(\rho_{0})}{\cos(\bar{\rho})-\coth
(\rho_{0})}\right)  ^{2}-\left(  2n^{2}-3+6\xi\right)  \left(  \frac{\cos
(\bar{\rho})\cosh(\rho_{0})-\sinh(\rho_{0})}{\cos(\bar{\rho})\sech(\rho
_{0})-\csch(\rho_{0})}\right) \nonumber\\
&  \ \ +\frac{\cosh^{2}(\rho_{0})\left(  4n^{2}-9+24\xi\right)  }{4}%
+\frac{\left(  4m^{2}+15\right)  }{4\sinh^{2}(\bar{\rho})}.
\end{align}

Even with the nonminimal coupling with the scalar curvature, in the extremal
values $\rho_{0}=0$ and $\rho_{0}\rightarrow\infty$ one also recovers shape
invariant potentials, since
\begin{align}
U_{\xi}^{\left(  0\right)  }\left(  \bar{\rho}\right)   &  =\left(  \frac
{15}{4}+m^{2}-20\xi\right)  \frac{1}{\sin^{2}\bar{\rho}}+n^{2}-\frac{9}%
{4}+6\xi+O\left(  \rho_{0}\right) \label{ceronmc}\\
U_{\xi}^{\left(  \infty\right)  }\left(  \bar{\rho}\right)   &  =\frac{1}%
{4}\frac{6(4\xi-1)\cos\bar{\rho}+4m^{2}+9-56\xi}{\sin^{2}\bar{\rho}}+O\left(
e^{-2\rho_{0}}\right)  \, ,\label{infnmc}%
\end{align}
corresponding to a Rosen-Morse potential and a Scarf potential, respectively.

It is interesting to note that for modes without angular momentum ($n=0$) and
$\xi=3/8$ both potentials lead to a fully resonant, equispaced spectra, which
might enhance the energy transfer between modes when nonlinearities are
included \cite{turbulentwormholes}, as it happens in AdS. Note that in a truly
quantum-mechanical problem nor the Rosen-Morse, neither the Scarf potential
lead to equispaced energies since in that case the eigenvalue is quadratic in
the principal quantum number, and actually the harmonic oscillator is the
unique potential with such property. Nevertheless, in our relativistic theory
the eigenvalue is quadratic in the frequencies, therefore all the potentials
which are quadratic in the principal quantum number, may lead to a equispaced
set of $\omega_{n}$ (see \cite{kleingordonization}).

The asymptotic behaviors for the function $R(z)$ with $z=(1-\tanh(\rho))/2$
that defined the radial dependence of the field \eqref{separacion}, is the same
as that given in equations \eqref{asymp} and \eqref{2}, replacing the,,
conformal weights by
\begin{equation}
\Delta^{\xi}_{\pm}=1\pm\frac{1}{2}\sqrt{m^{2}-20\xi+4}.
\end{equation}
Regarding the asymptotic behavior one can define an effective mass
$m_{\mathrm{eff}}^{2}:=m^{2}-20\xi$ and we will be focused in the region
$m_{\mathrm{eff}}^{2}\geq0$, in which the reflective boundary conditions lead
to a unique set of normal frequencies. Below we present the spectra for
different values of the parameters characterizing the potential, i.e.,
$(\xi,\rho_{0},m^{2},n)$.

\begin{figure}[h]
\includegraphics[scale=0.4]{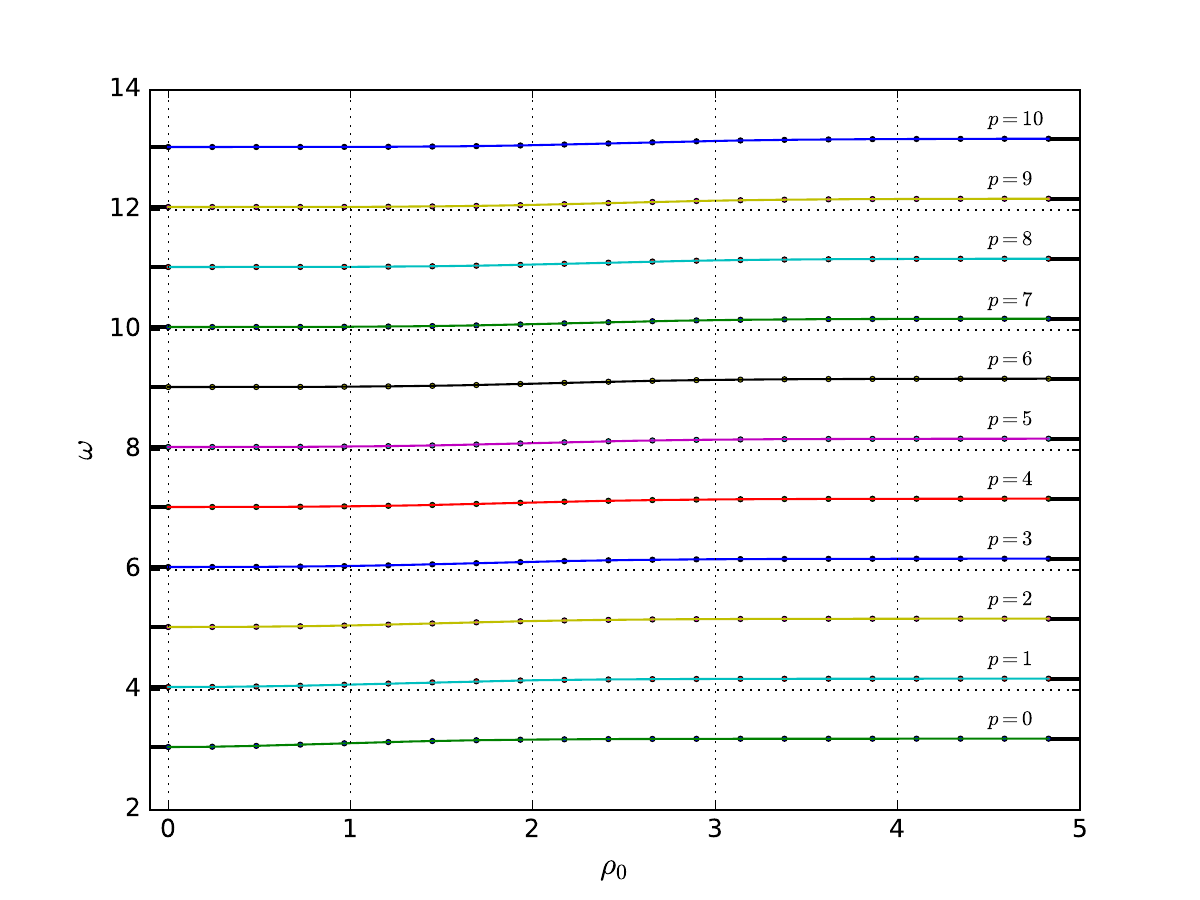}
\includegraphics[scale=0.4]{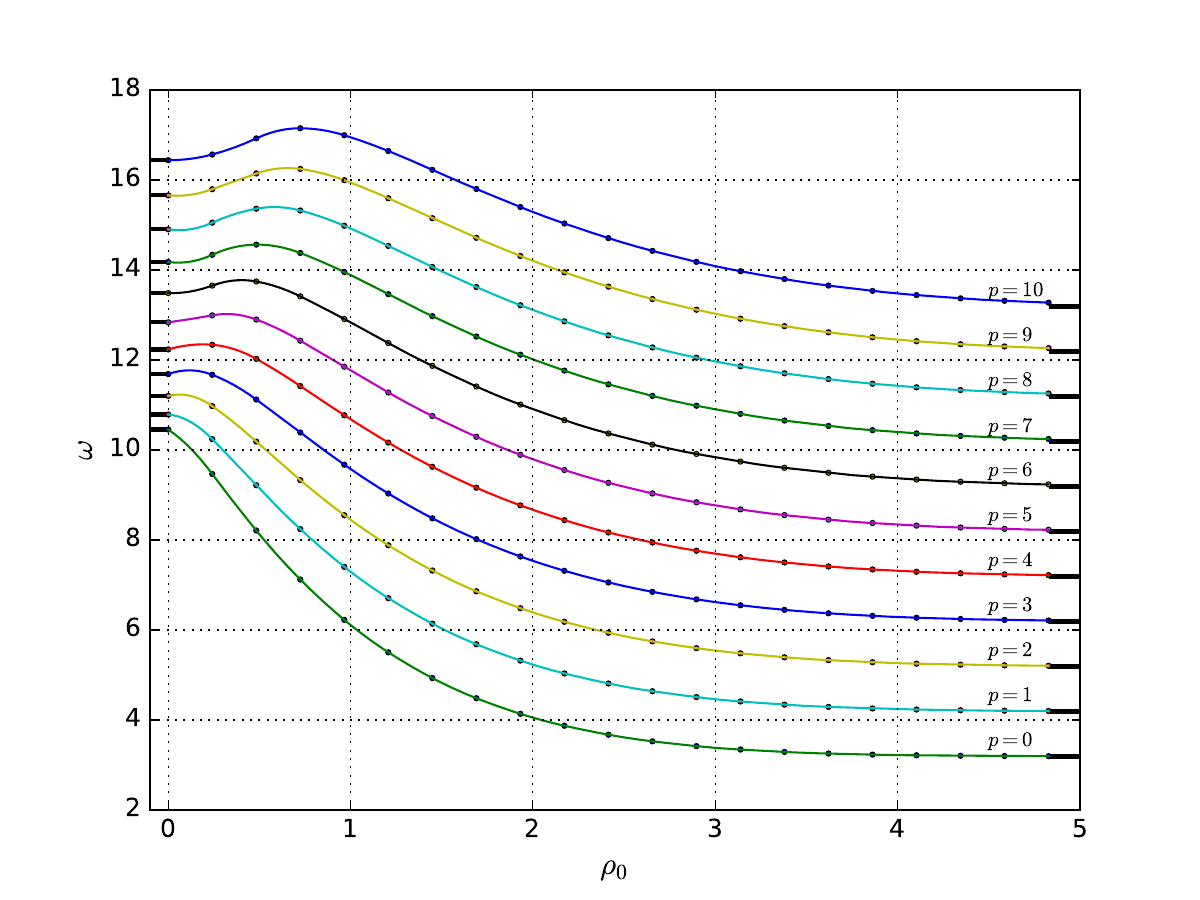}\caption{Fundamental and first
overtones for the nonminimally coupled scalar probe without angular momentum
($n=0$, left), and $n=10$ (right), for $m^{2}=10$ and $\xi=3/8$. The figure in
the left smoothly connects two fully resonant spectra for $\rho_{0}=0$ and
$\rho_{0}\rightarrow\infty$.}%
\label{n0n10xi0}%
\end{figure}

\begin{figure}[h]
\includegraphics[scale=0.4]{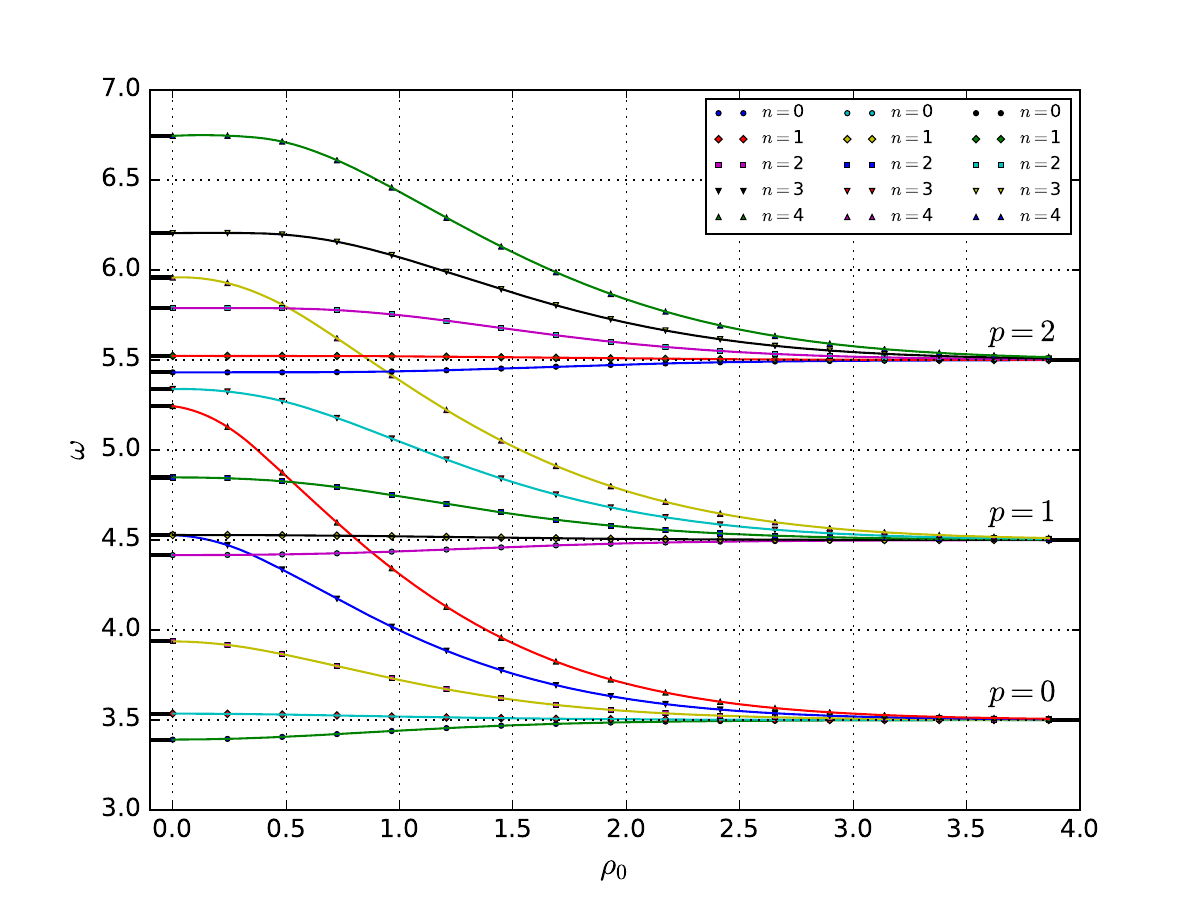}
\includegraphics[scale=0.4]{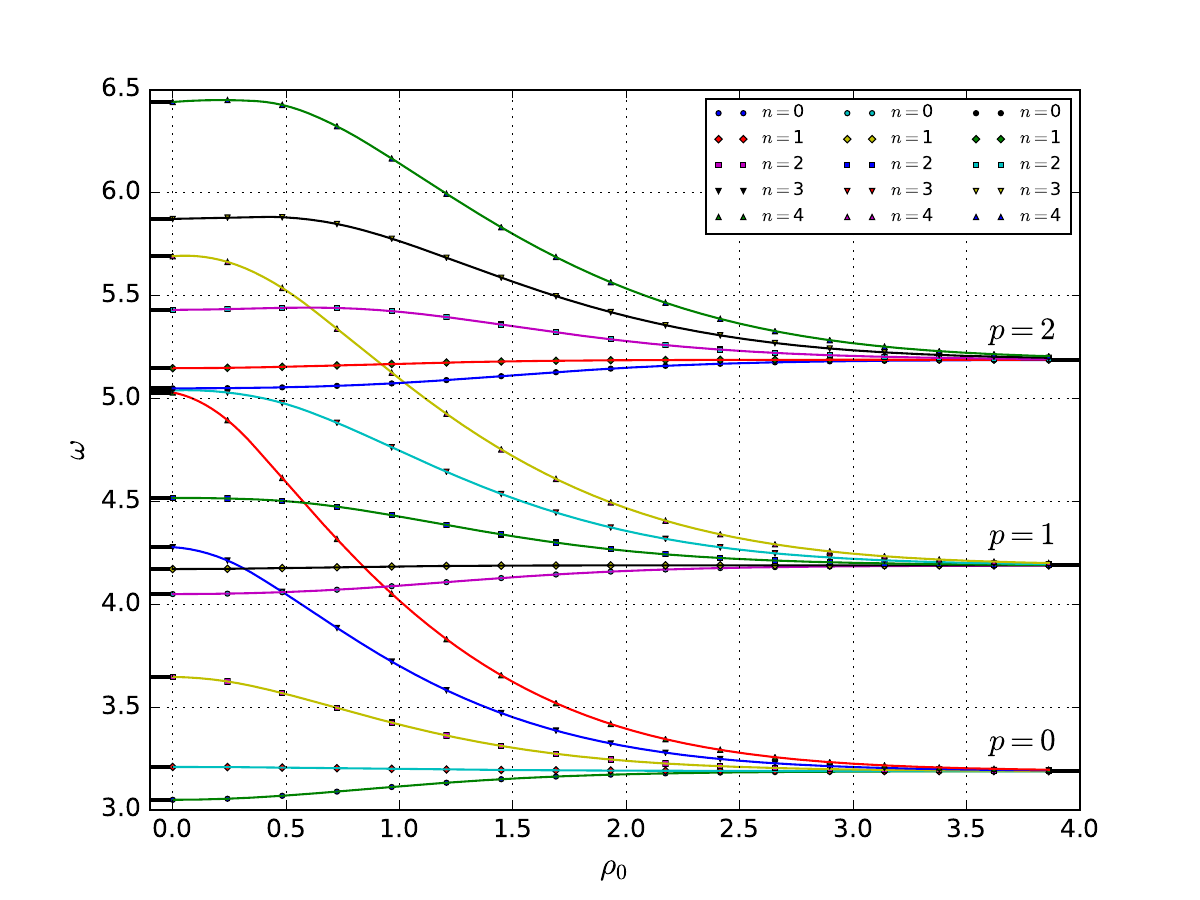}\caption{First three modes spectra for
$\xi=1/4$ (left) and $\xi=3/8$ (right) for $m^{2}=10$ for $n=0,1,2,3,4$.}%
\label{laultima}%
\end{figure}

\section{A new wormhole solution of Einstein-Gauss-Bonnet and Lovelock theories}
Before finishing, let us report on a new, wormhole solution of Einstein-Gauss-Bonnet or even Lovelock theories with a unique vacuum, provided \eqref{wormholed}, solve the corresponding field equations. The new wormhole geometry is given by
\begin{equation}
ds^{2}=l^{2}\left[  -\cosh^{2}\rho\   dt^{2}+d\rho
^{2}+\cosh^{2}\left(\rho-\rho_{0}\right)  d\phi^{2}+\cosh^2\rho \ d\Sigma_{2}^{2}  \right]
,\label{newwormholed}%
\end{equation}
that can be constructed from  \eqref{wormholed} via the double Wick rotation $t\rightarrow i\phi$ and $\phi\rightarrow it$. Notice that the double Wick rotation produces a different spacetime only when $\rho_0\neq 0$, which is exactly the case we are considering in the present work. The geodesic equation on this background, freezing the dynamics along the coordinates of $\Sigma_{2}$, which must be of constant Ricci scalar curvature equal to $-6$, leads to
\begin{align}
\dot{t}&=\frac{E}{l^2\cosh^2\rho}\ , \dot{\phi}=\frac{L}{l^2\cosh^2{\rho-\rho_0}}\\
l^2\dot\rho^2&=-b+\frac{E^2}{l^2\cosh^2\rho}-\frac{L^2}{\cosh^2\left(\rho-\rho_0\right)}\label{radialeq}
\end{align}
where $L$ and $E$, are the angular momentum and the energy of the particle and $b=+1,0,-1$ for timelike, null or spacelike geodesics, respectively. From the geodesic radial equation \eqref{radialeq}, one can see that the existence of a value of the radial coordinate $\rho=\rho_c$, for which the noncentrifugal (proportional to $E^2$) and the centrifugal (proportional to $L^2$) contributions balance, is nontrivial. In this case, the noncentrifugal contribution pushes toward the surface $\rho=0$, while the centrifugal contribution pushes particles away from the surface $\rho=\rho_0$, in consequence, in the region $0<\rho<\rho_0$, it is impossible to balance both contributions. A similar mechanism precludes the existence of orbits within the same region for the wormhole \eqref{wormholed} (see \cite{DOT1}). On the other hand, the equation for a scalar field probe \eqref{KG}, \eqref{separacion}, on the new geometry \eqref{newwormholed}, leads to the following equation for the radial dependence
\begin{align}
0 &=\partial_\rho\left[\cosh(\rho-\rho_0)\cosh^{3}(\rho)\partial_\rho R(\rho)\right]+\nonumber\\
& +\cosh(\rho-\rho_0)\cosh^{3}(\rho)\left[-\frac{n^2}{\cosh^2(\rho-\rho_0)}+\frac{\omega^2}{\cosh^2(\rho-\rho_0)}-m^2l^2\right]R(\rho) \label{ecuRrhonew}
\end{align}
which exactly corresponds to \eqref{ecuRrho} setting $d=5$, $\omega\rightarrow i n$ and $n\rightarrow i\omega$.

Before finishing this section, it is interesting to consider the energy content of the new wormhole geometry \eqref{newwormholed}. For Einstein-Gauss-Bonnet, in five dimensions, at the Chern-Simons point, the regularized action principle of \cite{olea} leads, via Noether theorem, to a definition for the energy of an asymptotically locally AdS spacetime. The mass of the new wormhole \eqref{newwormholed} receives contributions from both boundaries $\rho\rightarrow\pm\infty$, which as for the wormhole \eqref{wormholed} (see \cite{DOT1}), vanishes, since the contributions from both boundaries cancel each other, and in this case are given by
\begin{equation}
M^\text{new}_{\pm\infty}=\mp \frac{5\,\sigma_2}{8\pi G} \sinh{\rho_0}, 
\end{equation}
where $\sigma_2=2\pi \text{Vol}(\Sigma_2)$. In this normalization, the contribution to the mass coming from each boundary on the original wormhole geometry \eqref{wormhole} in dimension five reads $M^\text{old}_{\pm\infty}=\pm \frac{3\, \sigma_3}{8\pi G} \sinh{\rho_0}$. As mentioned above, in both cases the total mass vanishes.

\section{Conclusions}

In this work we have embedded in Lovelock theory in arbitrary dimensions, the $d=2n+1$-dimensional, wormhole geometry originally constructed in \cite{DOT1}. The selected theories in even dimension are characterized by possessing a unique, maximally symmetric AdS vacuum, and the geometry of the wormhole throat is restricted to fulfill a tensor \eqref{tensorconst} and a scalar \eqref{scalarconst} constraint, containing one curvature less than the original theory. We have also obtained the spectrum of normal modes for a scalar probe
on the asymptotically locally AdS wormhole in dimension five, complementing the work of \cite{Correa:2008nq}, where the case $\rho_0=0$ was considered, a case that can be solved in an analytic manner. Here we have shown that
the particular case with $\rho_0=0$ leads to a problem for the radial dependence of the scalar probe of the Schr\"{o}dinger type, in the well-known Rosen-Morse potential.
Remarkably, we have shown that when $\rho_{0}\rightarrow\infty$, the spectrum
can be also obtained in an analytic manner and it does not depend on the
angular momentum of the scalar. This exact result can be useful to obtain the
spectrum on a wormhole with a large, finite $\rho_{0}$ by perturbation
theory. As discussed in the Appendix the exact limit $\rho_{0}\rightarrow
\infty$ can be taken after suitable regularizations leading to a geometry that
also describes a wormhole. We left some of the details of such case to the
Appendix, since only one of the asymptotic regions is locally AdS in that setup.
It is important to mention that since the $g_{tt}$ component of the metric is
not a constant, the wormhole is not ultrastatic. Even more, for $\rho_{0}%
\neq0$ the minimum of the $g_{tt}$ component of the metric does not coincide
with the throat.

In \cite{Kim:2018ang}, the authors considered a scalar
probe propagating on the so-called natural wormholes that can be constructed
by smoothly matching spherically symmetric solutions of GR, coupled to
a Born-Infeld electrodynamics. In the effective radial eigenvalue problem, they imposed
reflective boundary condition at one of the asymptotic regions and ingoing
boundary conditions at the throat, leading to complex quasinormal
frequencies. In our case, we have defined our eigenvalue problem
by ensuring that the field is nondivergent at both asymptotic regions,
requiring reflective boundary conditions when $\rho\rightarrow\pm\infty$.

The frequencies of the scalar field are given with respect to the coordinate
time $t$, which in our case coincides with the proper time of a static
observer located at $\rho=\rho_{0}$. Note that such observer is a geodesic
one. In an asymptotically flat spacetime, as for example in the
Schwarzschild black hole, the frequencies are referred to an observer located
at the asymptotic region, which is independent of the value of the black hole
mass. Nevertheless, when a negative cosmological constant is included, as for
example in the four-dimensional Schwarzschild-AdS black hole, in
Schwarzschild-like coordinates, a time dependence of the form $e^{-i\omega t}$
in a scalar probe implies that the frequencies correspond to those measured by
an observer located at $r=\left( 2Ml^{2}\right) ^{1/3}$. Note that such
observer is nongeodesic, in contraposition to our geodesic observer measuring
frequencies in the wormhole. From the holographic viewpoint this coordinate
time is actually the time in the dual CFT.

We have also included a nonminimal coupling between the scalar probe and the
scalar curvature, which allows, for some particular values, to connect two fully resonant, equispaced spectra. It is worth to mention here also that
fully resonant, equispaced spectra play a central role in the turbulent energy
transfer that leads to nonperturbative AdS instability, see, e.g.,
\cite{Turbulent1,Dias:2011ss,Turbulent3,Craps:2014vaa,Craps:2014jwa,Turbulent6}. This interesting phenomenology has also
been observed in other nonlinear models as gravitating scalars on a spherical
cavity in 3+1 \cite{Maliborski:2012gx}, on systems governed by the
Gross-Pitaevskii equation \cite{Biasi:2018iiu}, on conformal dynamics on the
Einstein Universe \cite{Bizon:2016uyg} and on vortex precession in
Bose-Einstein condensates \cite{Biasi:2017pdp}.

Finally, we have also introduced a new family of wormhole geometries, that are obtained from the first set via a double Wick rotation. The propagation of a test particle on this new geometry, unveils a region where no geodesic, circular orbit exist, and the spectrum of a scalar field probe can also be obtained from the one on the seed geometries by a suitable Wick rotation in frequency/momentum domain.

It would be interesting to explore the sector with negative squared masses.
Since there is an asymptotically locally AdS asymptotic behavior, it is
natural to expect the existence of an effective Breitenlohner-Freedman bound
that may depend on $\rho_{0}$ and that would have to be obtained numerically.
We expect to report on this problem in the future.

\section*{Acknowledgments}

{The authors want to thank Eloy-Ayon Beato, Francisco Correa, Danilo Diaz,
Oleg Evnin and Nicolas Grandi for enlightening comments. This work was
partially supported by CONICYT Fellowships 21161099 and 22171243, FONDECYT Grants No. 1181047, No. 11191175, No. 1221504, No. 1200022, No. 1200293, No. 1210500, No. 1210635.
%O.F. is supported by the Project DINREG 19/2018 of the
% Direcci\~{A}%
% %TCIMACRO{\U{b3}}%
% %BeginExpansion
% ${{}^3}$%
% %EndExpansion
% n de Investigaci\~{A}%
% %TCIMACRO{\U{b3}}%
% %BeginExpansion
% ${{}^3}$%
% %EndExpansion
% n of the Universidad Cat\~{A}%
% %TCIMACRO{\U{b3}}%
% %BeginExpansion
% ${{}^3}$%
% %EndExpansion
% lica de la Sant\~{A}-sima Concepci\~{A}%
% %TCIMACRO{\U{b3}}%
% %BeginExpansion
% ${{}^3}$%
% %EndExpansion
% n.}

\section*{APPENDIX}

Here we discuss some features of the scalar probe on the spacetime obtained
after a suitable regularization in the limit $\rho_{0}\rightarrow+\infty$. It
is useful to return to Schwarzschild-like coordinates, that cover part of the
wormhole spacetime (\ref{wormhole}). In such coordinates the metric reads
\cite{DOT2}
\begin{equation}
ds^{2} = -\left( \frac{r}{l^{2}} + a\sqrt{\frac{r^{2}}{l^{2}}-1} \right) ^{2}
d \bar{t}^{2} + \left( \frac{r^{2}}{l^{2}} -1 \right) ^{-1} dr^{2} + r^{2}
(d\phi^{2} + d\Sigma_{2}^{2}) ,\label{whsolution}%
\end{equation}
where $r = l \cosh(\rho)$, $t =\bar{t}/(l\cosh(\rho_{0})) $ and the
integration constant $\rho_{0}$ relates to $a$ by $\rho_{0}:=-\tanh^{-1}(a)$.
Now we consider that $\rho_{0} \rightarrow\infty$. In this case $a=-1$ and
returning to the proper radial coordinate $\rho$ one obtains
\begin{equation}
ds^{2} = l^{2} \left[ - e^{-2\rho} dt^{2} + d\rho^{2} + \cosh^{2}(\rho)
(d\phi^{2} + d\Sigma_{2}^{2}) \right]  ,\label{whrhoceroinf}%
\end{equation}
where $t=\bar{t}/l $. This solution also describes a wormhole geometry with a
traversable throat located at $\rho=0$. Note that this spacetime is
asymptotically locally AdS only when $\rho\rightarrow-\infty$. Hereafter we
set $l=1$.

The radial equation for the nonminimally coupled scalar probe on this
wormhole, for the ansatz $\Phi=e^{-i \omega t} e^{i n \phi} R(\rho)$ reduces
to
\begin{equation}
\label{eqRrho}\frac{d^{2}}{d\rho^{2}} R(\rho) - (1-3\tanh(\rho)) \frac
{d}{d\rho} R(\rho) + \left(  e^{2 \rho} \omega^{2} -m^{2} + (14 - 6 \tanh
(\rho)) \xi- \frac{n^{2}}{\cosh^{2}(\rho)} \right)  R(\rho) = 0.
\end{equation}
The asymptotic behavior of the solution is given by
\begin{align}
R(\rho)  & \overset{\rho\rightarrow-\infty}{\sim} A_{1} \ e^{(2-\sqrt
{4+m^{2}-20\xi})\rho} + A_{2} \ e^{(2+\sqrt{4+m^{2}-20\xi})\rho} +
\mathcal{O}(e^{2\rho}),\\
R(\rho)  & \overset{\rho\rightarrow\infty}{\sim} B_{1} \ e^{-3\rho/2-i\omega
e^{\rho}} + B_{2} \ e^{-3\rho/2+i\omega e^{\rho}} + \mathcal{O}(e^{-\rho}).
\end{align}
As expected the behavior of the scalar when $\rho\rightarrow-\infty$ is a
power law behavior on the areal coordinate $r\sim e^{\rho}$, typical of
scalars on asymptotically AdS regions. While the behavior at the other non-AdS
asymptotic region corresponds to that of an ingoing and outgoing wave. The
equation can be solved analytically in terms on confluent Heun functions
\cite{Heun}. The equation can be recast in a Schr\"{o}dinger form by
scaling the radial function and considering the ansatz $\Phi=e^{-i \omega t}
e^{i n \phi} u(\rho)/\cosh^{3/2}(\rho)$ and using the new radial coordinate
$\bar{\rho}=e^{\rho}$, with $\bar{\rho} \in(0,\infty)$, leading to
\begin{equation}
\label{eqRschrodinger}-\frac{d^{2}}{d\bar{\rho}^{2}} u(\bar{\rho}) + U
u(\bar{\rho})= \omega^{2} u(\bar{\rho}) ,
\end{equation}
with the effective potential explicitly given in terms of the coordinate
$\bar\rho$ by
\begin{equation}
U:=\frac{m^{2}}{\bar{\rho}^{2}}-\frac{4(2\bar{\rho}^{2}+5)}{\bar{\rho}%
^{2}(\bar{\rho}^{2}+1)}\xi+\frac{4 n^{2}}{(\bar{\rho}^{2}+1)^{2}}+\frac{3}{4}
\frac{\bar{\rho}^{4} + 2\bar{\rho}^{2} + 5}{(\bar{\rho}^{2}+1)^{2} \bar{\rho
}^{2}}.
\end{equation}

Note that this potential depends explicitly on the angular momentum of the
scalar $n$ while $U^{(\infty)}_{\xi}$ defined in \eqref{infnmc}, does not.
This confirms what we previously discuss in the paper regarding the fact that
the exactly solvable potential obtained for the leading term of
\eqref{infnmc}, has only to be thought as a tool to perturbatively obtain the
frequencies for large $\rho_{0}$.

Note that here, at least on one side of the wormhole it is clear how to
recognize ingoing and outgoing modes. It would be interesting in this case to
compute the transmission and reflection coefficients of the wormhole,
along the lines of the computation of graybody factors in
asymptotically AdS black holes \cite{GBF}.

\end{document}